\documentclass[prd,amsmath,amssymb,superscriptaddress,nofootinbib]{revtex4}

\usepackage{graphicx}
\usepackage{dcolumn}
\usepackage{bm}

\normalbaselines \topmargin -0.25 truein \oddsidemargin 0.30
truein \evensidemargin 0.30 truein 

\begin{document}

\title{Einstein-Yang-Mills-aether theory with nonlinear axion field:  \\ Decay of color aether and the axionic dark matter production}

\author{Alexander B. Balakin}
\email{Alexander.Balakin@kpfu.ru} \affiliation{Department of General
Relativity and Gravitation, Institute of Physics, Kazan Federal University, Kremlevskaya
str. 18, Kazan 420008, Russia}
\author{Gleb B. Kiselev}
\email{gleb@karnaval.su} \affiliation{Department of General
Relativity and Gravitation, Institute of Physics, Kazan Federal University, Kremlevskaya
str. 18, Kazan 420008, Russia}

\date{\today}

\begin{abstract}
We establish a nonlinear  version of the SU(N) symmetric theory, which  describes self-consistently the interaction between the gravitational, gauge, vector and pseudoscalar (axion) fields. In the context of this theory the SU(N) symmetric multiplet of vector fields is associated with the color aether, the decay of which in the early Universe produced the canonic dynamic aether and the axionic dark matter. The SU(N) symmetric Yang-Mills field, associated with the color aether, forms the source, which transfers the energy of the decaying color aether to the axion field.
The nonlinear modification of the model uses explicitly the requirement of discrete symmetry, prescribed by the axion field, and is based on the analogy with nonlinear physical pendulum. We show that in the framework of this nonlinear regular model the axion field can grow to an arbitrarily large value, thus explaining the abundance of the axionic dark matter in the Universe.

\end{abstract}

\maketitle

\section{Introduction}\label{Intro}

Dark matter, the important element of the Universe structure, highly likely consists of axions and axion-like massive particles \cite{DM1,DM2,DM3,DM4,DM5,DM6,DM7}. There is a number of scenaria according to which the relic axions could be produced in the early Universe \cite{a1,a2,a3,a4,a5,a6,a7,a8,a9,a10,a11,a12,a13,a14,a15,a16}. One of these scenaria is connected with the idea of contact coupling between the gauge  and pseudoscalar fields.
This idea arises naturally, when one sees the interaction term $\frac14 g_A \theta \ {}^*\! F^{(a)}_{mn}F^{mn}_{(a)}$ in the Lagrangian discussed in the classical works \cite{PQ,W,F}. The term equal to the convolution of the Yang-Mills field strength $F^{mn}_{(a)}$ with its dual ${}^*\! F^{(a)}_{mn}$ multiplied by the pseudoscalar field $\theta$ and sealed by the coupling constant $g_A$,  indicates unambiguously that the evolutionary equation for the axion field itself has to have the pseudo-source $-\frac14 g_A {}^*\! F^{(a)}_{mn}F^{mn}_{(a)}$ formed by the SU(N) symmetric gauge field. In other words, when the Yang-Mills field configuration provides this pseudo-source to be non-vanishing, we have the right to expect that the axion field can be produced due to such interaction. But there are three questions in this reasoning.

I. The first question is connected with the origin of the SU(N) symmetric gauge field. The multiplet of gauge field potentials $A^{(a)}_m$ can appear in the SU(N) symmetric theory as the element of the extended covariant derivative acting on the scalar, vector, spinor, etc. fields.
For instance, when  we work with the multiplet of vector fields $U^{j(a)}$, the usage of the extended covariant derivative of the vector field multiplet
$
D_k U^{j(a)}= \nabla_k U^{j(a)} {+} g f^{(a)}_{ \ (b)(c)} A^{(b)}_k U^{j(c)}
$
guarantees the theory to possess the SU(N) invariance, on the one hand, and provides the gauge field to become the inherent element of the theory, on the other hand. Thus, one can consider the following scenario of the axion production... Let us imagine that in the early Universe we deal with the SU(N) symmetric  multiplet of vector fields  indicated in the works \cite{color1,color2} as the SU(N) symmetric dynamic aether, or the color aether, for short. The model of the color aether is considered to be a SU(N) extension of the model of the dynamic aether, introduced by Jacobson with colleagues (see, e.g., \cite{J1,J2,J3,J4,J5}; this color aether converts into the dynamic aether due to the spontaneous color polarization \cite{color2}.  So if the gauge field, associated with the color aether, interacts with the pseudoscalar field, we could obtain the axion production, and these relic axions could form now the axionic dark matter.

II. The second question is connected with the energy transfer from the gauge field to the axion configuration. Clearly, a huge number of axions should be born as a result of interactions between the gauge and pseudoscalar fields in the early Universe, if now these particles form the dark matter, associated with 23\% of the Universe energy density. In other words, we assume that in the course of decay the energy of the color aether was partially transferred to the canonic dynamic aether, and partially to the axion configuration, the gauge field being the mediator of this process.  We assume that the theory, which could explain this phenomenon, has to be nonlinear (see, e.g., \cite{Non0,Non1,Non2,Non3} for motivation and examples of modeling in the U(1) symmetric theory).

III. The third question concerns the symmetry of the model, since it is the symmetry that guarantees the rigor of the theory formulations in the absence of a sufficient amount of observational data. In the works \cite{Non2,Non3}, which deal with nonlinear axion electrodynamics, we considered two special internal symmetries in addition to the standard requirements of general relativistic approach and in addition to the standard U(1) symmetry of electrodynamics. First, we took into account that the nonlinear model has to be invariant with respect to the discrete transformation of the axion field $\theta \to \theta {+} 2\pi n$ ($n$ is an integer). Second, we assumed that the axion electrodynamics also is invariant with respect to the Jackson's symmetry, which also can be indicated as the SO(2) symmetry of classical electrodynamics \cite{Jackson}. This symmetry assumes that the Faraday-Maxwell equations hold the form, if we use the linear transformations of the electric and magnetic field four-vectors  $E^k \to {\tilde{E}^k} = \cos{\varphi} E^k {+} \sin{\varphi} B^k $ and $B^k \to {\tilde{B}^k} = {-} \sin{\varphi} E^k {+} \cos{\varphi} B^k $ with some constant angle $\varphi$. When we deal with the axion chromodynamics (SU(3) symmetric theory), or with general SU(N) symmetric models involving the Yang-Mills fields coupled to the axion field, of course, we have to use again the discrete symmetry $\theta \to \theta {+} 2\pi n$. As for the Jackson's symmetry, its generalization for the Yang-Mills theory bumps into one disputable moment. The Yang-Mills field strength tensor $F^{(a)}_{mn}=\nabla_m A^{(a)}_n {-} \nabla_n A^{(a)}_m {+} g f^{(a)}_{\ (b)(c)} A^{(b)}_m A^{(c)}_n$ with the group index $(a)$ contains not only the potential $A^{(a)}_m$ with the same group index, but all other vector potentials also, if the SU(N) group constants $f^{(a)}_{\ (b)(c)}$ are non-vanishing. In other words, for the non-Abelian Yang-Mills fields the formal application of the Jackson's symmetry seems to be not sufficiently justified.
That is why, in this work we suggest to use the idea of generalization, which is associated with the model of physical pendulum in the classical mechanics. One can recall, that the equation $\ddot{x}{+} \omega^2_0 \sin{x}=0$, which  describes  the physical pendulum, converts into the equation of harmonic oscillations of the mathematical pendulum $\ddot{x} {+} \omega^2_0 x =0$ ,  when $x$ is small. For the physical pendulum the constant $\omega_0$ can be eliminated by the time transformation $t \to \tau = \omega_0 t$, i.e., the parameter $T_0 = \frac{1}{\omega_0}$ defines the typical time scale of the evolution; in the case of mathematical pendulum $\omega_0$ relates to the frequency of the harmonic oscillations. Mention should be made that formally speaking the equation of physical pendulum is invariant with respect to discrete transformation $x \to x+ 2\pi n$ ($n$ is an integer), while the equation of the mathematical pendulum does not possesses this symmetry. As the result, the solutions to the second equation remain finite, however, the equation of the physical pendulum admits the solutions, which grow with time quasi-periodically (see, e.g., \cite{Arnold} for details).

Keeping in mind the analogy with physical pendulum and taking into account the discrete symmetry, which the theory of axions insists on, we consider below the nonlinear Lagrangian of the Yang-Mills field in the form ${\cal L}({\cal I})$ with the unified invariant ${\cal I}$ given by
\begin{equation}
{\cal I} = \frac14 \left[ F^{(a)}_{mn} F_{(a)}^{mn} + \sin{\phi} {}^*\!F^{(a)}_{mn} F_{(a)}^{mn}\right] \,.
\label{001}
\end{equation}
Here $\phi$ denotes the dimensionless pseudoscalar field associated with the axionic dark matter. The function ${\cal L}({\cal I})$ is assumed to have at least two continuous derivatives and tends to ${\cal I}$ for small argument.
The color aether coupled to the axion field, is presented below using the extended constitutive tensor, nonlinear in the axion field, which also supports the discrete symmetry prescribed by the axion field.

To conclude, we would like to emphasize that the novelty of this work is associated with three items. First, based on the nonlinear version of the axion electrodynamics constructed in \cite{Non2,Non3}, we develop the generalized nonlinear axion chromodynamics, i.e., we use the SU(N) symmetric Yang-Mills field instead of U(1) symmetric electromagnetic field. Second, we assume that the configuration of coupled gauge and pseudoscalar fields possesses the symmetry indicated as symmetry of physical pendulum, which came to replace the Jackson's symmetry of the nonlinear axion electrodynamics. Third, we consider the backreaction of the axion field on the color aether, which is encoded in the structure of the generalized Jacobson's constitutive tensor; in this sense, the program of axionic extension of the dynamic aether fulfilled in \cite{AA} is now (partially) expanded on the model of color aether with gauge field.

The paper is organized as follows. In Section II the formalism of the theory is presented: starting from the nonlinearly extended Lagrangian of the coupled SU(N) symmetric quartet, which includes the gravitational, gauge, vector and pseudoscalar fields, the self-consistent set of master equations is derived. In Section III based on the idea of the spontaneous color polarization, we reduced the master equations for the case of vector and gauge fields parallel in the group space, then we considered the application of this truncated model to the anisotropic cosmological model of the Bianchi-I type, and have found the exact solutions to the equations for the gauge and vector fields. In Section IV we focus on the solutions to the equation for the axion field, and discuss the possibilities for the axion field growth. Section V contains conclusions.

\section{The formalism}

\subsection{Action functional and basic definitions}

The details of interaction between gravitational, gauge, vector and pseudoscalar fields are encoded in the action functional
$$
S {=} \int d^4 x \sqrt{{-}g} \left\{ \frac{1}{2\kappa}\left[R {+} 2\Lambda {+} \lambda \left(U^m_{(a)}U^{(a)}_m {-}1 \right) {+} {\cal K}^{ijmn}_{(a)(b)}D_i U^{(a)}_{m}D_j
U^{(b)}_n \right] + \right.
$$
\begin{equation}
\label{SUNact}
 \left. {+} {\cal L}({\cal I}) {+} \frac12 \Psi^2_0  \left[V(\phi) {-} \nabla_k \phi \nabla^k \phi \right] \right\} \,.
\end{equation}
The gravitational field is described using the standard Einstein-Hilbert Lagrangian: $\sqrt{-g}$ is constructed using the determinant of the spacetime metric $g_{mn}$; $R$ is the Ricci scalar, $\Lambda$ is the cosmological constant, $\kappa {=} 8 \pi G$ is the Einstein constant ($c{=}1$). Other elements of the total Lagrangian require more detailed definitions.

\subsubsection{Multiplet  of vector fields}

We work with the  multiplet of real vector fields $U^{j(a)}$, which appear in the adjoint representation of the SU(N) group \cite{Rubakov}. We assume that the vector fields satisfy the condition
$U^{m(a)}U^{n(b)} g_{mn} G_{(a)(b)} {=}1$. The group index $a$ runs in the range $a=1,2,...N^2{-}1$. $G_{(a)(b)}$ is the metric in the group space; it is defined as follows
\begin{equation}
G_{(a)(b)} = \left( {\bf t}_{(a)} , {\bf t}_{(b)} \right) \equiv 2 {\rm Tr} \
{\bf t}_{(a)} {\bf t}_{(b)}
\label{scalarproduct}
\end{equation}
via the SU(N) group generators ${\bf t}_{(a)}$, which satisfy the commutation relations
\begin{equation}
\left[ {\bf t}_{(a)} , {\bf t}_{(b)} \right] = i  f^{(c)}_{\ (a)(b)} {\bf t}_{(c)} \,. \label{fabc}
\end{equation}
This formula involves the SU(N) group constants $f^{(a)}_{\ (b)(c)}$, which satisfy the Jacobi identity
\begin{equation}
f^{(a)}_{\ (b)(c)} f^{(c)}_{\ (e)(h)}  {+} f^{(a)}_{\ (e)(c)} f^{(c)}_{\ (h)(b)} {+} f^{(a)}_{\ (h)(c)} f^{(c)}_{\ (b)(e)} {=} 0 \,. \label{jfabc}
\end{equation}
The structure constants with three subscripts
\begin{equation}
f_{(c)(a)(b)} \equiv G_{(c)(d)} f^{(d)}_{\ (a)(b)} =
 {-} 2 i
{\rm Tr}  \left[ {\bf t}_{(a)},{\bf t}_{(b)} \right] {\bf
t}_{(c)}   \label{fabc1}
\end{equation}
are antisymmetric with respect to transposition of any two indices.
The Lagrange multiplier $\lambda$ appeared in front of the term $\left(U^m_{(a)}U^{(a)}_m {-}1 \right)$ stands to guarantee that this term is equal to zero.
The SU(N) symmetric extended covariant derivative \cite{Akhiezer} in the application to the vector field yields
\begin{equation}
D_k U^{j(a)} = \nabla_k U^{j(a)} + g f^{(a)}_{ \ (b)(c)} A^{(b)}_k U^{j(c)} \,,
\label{DU}
\end{equation}
where $\nabla_k$ is the spacetime covariant derivative.
It includes the coupling constant $g$ and the multiplet of co-vector potentials $A^{(a)}_m$, describing the gauge field  attributed to the SU(N) group.
The metric $G_{(a)(b)}$ and the structure constants $f^{(d)}_{\ (a)(c)}$ satisfy the conditions
\begin{equation}
D_m G_{(a)(b)} = 0 \,, \qquad
D_m f^{(a)}_{\ (b)(c)} = 0 \,, \label{DfG}
\end{equation}
i.e., they are the gauge covariant constant tensors in the group space:

\subsubsection{Yang-Mills tensors}

The multiplet of tensors $F^{(a)}_{mn}$ is expressed via $A^{(a)}_m$ as follows:
\begin{equation}
F^{(a)}_{mn} = \nabla_m
A^{(a)}_n - \nabla_n A^{(a)}_m + g f^{(a)}_{\ (b)(c)}
A^{(b)}_m A^{(c)}_n \,. \label{46Fmn}
\end{equation}
This multiplet of tensors describes the gauge field strength and is indicated by the term Yang-Mills tensors.
The dual tensors ${}^*\! F^{ik(a)}$
\begin{equation}
{}^*\! F^{ik(a)} = \frac{1}{2}\epsilon^{ikls} F^{(a)}_{ls} \,,
\label{dual}
\end{equation}
are based on the universal Levi-Civita pseudo tensor $\epsilon^{ikls}= \frac{E^{ikls}}{\sqrt{-g}}$ ($E^{0123}=1$). When we use the definition (\ref{46Fmn}), we obtain the identity
\begin{equation}
D_k {}^*\! F^{ik(a)} \equiv \nabla_k  \ ^*\! F^{ik(a)} +
g f^{(a)}_{\ (b)(c)}
A^{(b)}_k  {^*\! F^{ik(c)}} = 0 \,.
\label{dual002}
\end{equation}

\subsubsection{Axion field}

The pseudoscalar (axion) field is described by the term $\phi$ (here we use the dimensionless quantity $\phi$ instead of $\theta$). $\Psi_0$ is the coupling constant describing the interaction of the axions with the gauge field. According to the theory of axions, the Lagrangian of the axionically active system has to be invariant with respect to discrete symmetry $\phi \to \phi + 2 \pi n $ ($n$ is an integer), that is why we consider the potential of the axion field  $V(\phi)$ to be of the periodic form
\begin{equation}
V(\phi) = 2 m^2_A \left(1-\cos{\phi} \right) \,.
\label{14}
\end{equation}
The term $\frac{1}{4} \sin{\phi} {}^*\!F^{(a)}_{mn} F^{mn}_{(a)}$ in the unified invariant (\ref{001}) also satisfies this discrete symmetry. As well as, the function $\sin{\phi}$ is odd, thus it changes the sign, when $t \to -t$, as the properties of the pseudoscalar field require. When $\phi \to 0$ the mentioned term converts into  $\frac{1}{4} \phi {}^*\!F^{(a)}_{mn} F^{mn}_{(a)}$ recovering the original term introduced by Peccei and Quinn \cite{PQ}.

\subsubsection{Extended Jacobson's constitutive tensor}

The structure of the constitutive tensor
\begin{equation}\label{constitutive1}
{\cal K}^{ijmn}_{(a)(b)} =  \cos{\phi}\left[G_{(a)(b)} \left(C_1 g^{ij} g^{mn} {+} C_2 g^{im}g^{jn}
{+} C_3 g^{in}g^{jm}  \right) + C_{4} U^{i}_{(a)} U^{j}_{(b)} g^{mn} \right]+
\end{equation}
$$
 {+} \sin{\phi}\left\{C_5  G_{(a)(b)} \epsilon^{ijmn}  {+} C_6  \left[U_{s(a)} \left(U^i_{(b)} \epsilon^{jsmn} {+} U^j_{(b)} \epsilon^{isnm} \right) {+} U_{s(b)} \left(U^i_{(a)} \epsilon^{jsmn} {+} U^j_{(a)} \epsilon^{isnm} \right)\right] \right\} \,,
$$
which we use in the Lagrangian, can be explained as follows. First, when $\phi=0$, and there is only one unit vector field $U^i$, this term recovers the Jacobson's constitutive tensor  \cite{J1}. Second, this tensor holds the form when we use the transformation $i \to j$, $m \to n$, $a \to b$ simultaneously, i.e., it supports the internal symmetry of the kinetic term ${\cal K}^{ijmn}_{(a)(b)}D_i U^{(a)}_{m}D_j U^{(b)}_n $. Third, this constitutive tensor includes $\phi$ as the argument of the trigonometric functions only, thus supporting the requirement of the discrete symmetry, prescribed by the axion field. Clearly, without the pseudoscalar field, i.e., when $\phi=0$, the Jacobson's constitutive tensor could not contain the Levi-Civita pseudo-tensor $\epsilon^{ijmn}$. In other words, axionic extension of the Einstein-Yang-Mills-aether theory allows us to add principally new terms and new phenomenological constants $C_5$ and $C_6$. As it was mentioned in \cite{color1} there are also a lot of possibilities to introduce new coupling constants, however, we omit them, since after the decay of the color aether, when all the vector fields become parallel in the group space, the mentioned terms disappear and the corresponding coupling constants become hidden. To conclude, in addition to the coupling constants $C_1$, $C_2$, $C_3$ and $C_4$ introduced in the classical work \cite{J1}, we consider two new coupling constants $C_5$ and $C_6$, which are appropriate just for the axionic extension of the  theory.

\subsection{Master equations of the model}

The total set of model equations can be obtained by the variation with respect to gauge potentials $A^{(a)}_i$, Lagrange multiplier $\lambda$, vector fields $U^{j(a)}$, pseudoscalar field $\phi$, and the metric $g^{pq}$. This procedure gives the following results.

\subsubsection{Equations for the Yang-Mills fields}

Variation of the action functional (\ref{SUNact}) with respect to $A^{(a)}_i$ gives the extended Yang-Mills equations
\begin{equation}\label{0Col1}
D_k \left\{{\cal L}^{\prime}({\cal I}) \left[F^{ik}_{(a)} + \sin{\phi} \ {}^*\! F^{ik}_{(a)} \right] \right\} = \Gamma^i_{(a)}\,,
\end{equation}
where the prime denotes the derivative with respect to the argument of the function. The color current $\Gamma^i_{(a)}$
\begin{equation}\label{Col4}
 \Gamma^i_{(a)} =  \frac{g}{\kappa}   f^{(d)}_{\ (c)(a)} U^{(c)}_m  {\cal K}^{ijmn}_{(d)(b)} D_j U^{(b)}_n
\end{equation}
appears because of variation of the second term in the right-hand side of (\ref{DU}) with respect to $A_i^{(a)}$.
If we deal with the quasi-linear model, i.e., ${\cal L} = {\cal I}$, keeping in mind (\ref{dual002}) we can rewrite (\ref{0Col1}) in the form
$$
\nabla_k F^{ik}_{(a)} + g f^{(c)}_{\ (b)(a)} A^{(b)}_k F^{ik}_{(c)}  =
$$
\begin{equation}\label{Col1}
= -   {}^*\! F^{ik}_{(a)} \cos{\phi} \nabla_k \phi + \frac{g}{\kappa}   f^{(d)}_{\ (c)(a)} U^{(c)}_m  {\cal K}^{ijmn}_{(d)(b)} \left(\nabla_j U^{(b)}_n + g f^{(b)}_{\ (h)(e)} A^{(h)}_j U^{(e)}_n \right)\,.
\end{equation}
The first term in the right-hand side of this equation for the gauge field  can be interpreted as the source associated with the axion field, and the second term relates to the contribution of the color aether. In other words, we deal with two independent sources, which prescribe the details of  the Yang-Mills field evolution.

\subsubsection{Equations for the vector fields}

Variation of the action functional (\ref{SUNact}) with respect to the Lagrange multiplier $\lambda$ gives the normalization condition $U^k_{(a)}U_k^{(a)}=1$.
Variation  with respect to the vector fields $U^{(a)}_m$ yields the balance equation
\begin{equation}
D_i {\cal J}^{im}_{(a)}
 = \lambda \ U^m_{(a)}  +  {\cal I}^{m}_{(a)} \,.
\label{CU1}
\end{equation}
Here the tensors ${\cal J}^{im}_{(a)}$ can be presented in the form
\begin{equation}
{\cal J}^{im}_{(a)} \equiv {\cal K}^{ijmn}_{(a)(b)} D_j U^{(b)}_n =
\label{CU2}
\end{equation}
$$
=\cos{\phi}\left[C_1 D^i U^{m}_{(a)}  {+} C_2 g^{im} D_n U^{n}_{(a)}
{+} C_3 D^m U^{i}_{(a)} + C_{4} U^{i}_{(a)} U^{j}_{(b)} D_j U^{m (b)} \right] {-} 2 C_5  \sin{\phi} \ \Omega^{im}_{(a)} -
$$
$$
{-} C_6  \sin{\phi} \left[ 2 \Omega^{ms(b)}\left(U_{s(a)} U^i_{(b)} + U_{s(b)} U^i_{(a)}\right) \
+ \epsilon^{ismn}D_j U_{n}^{(b)} \left(U_{s(a)} U^j_{(b)} + U_{s(b)} U^j_{(a)}  \right) \right] \,.
$$
In this formula we used the new definition for the multiplet of pseudotensors
\begin{equation}
\Omega^{lm(b)} \equiv \frac12 \epsilon^{jlmn} D_j U_n^{(b)} \,,
\label{CU319}
\end{equation}
which are skew-symmetric with respect to the indices $l$ and $m$, and play the roles of color vortex operators in our context.
The term ${\cal I}^{m}_{(a)}$ in (\ref{CU1}) appeared due to variation of the extended constitutive tensor with respect to the vector fields; its structure is the following:
$$
{\cal I}^l_{(d)} \equiv \frac12  D_i U_{m}^{(a)} D_j U_{n}^{(b)} \frac{\delta}{\delta U^{(d)}_l} \ {\cal K}^{ijmn}_{(a)(b)}  =  C_{4} \cos{\phi} D^l U^{n}_{(d)} \  U^{j}_{(b)}  D_j U_{n}^{(b)}  +
$$
\begin{equation}
 + 2C_6  \sin{\phi} \left[ \Omega^{lm(b)} U^p_{(b)} D_p U_{m(d)} {+} \Omega^{lm}_{(d)} U^p_{(b)} D_p U_{m}^{(b)} +
 U_{s(b)} \left(\Omega^{sm}_{(d)} D^l U_m^{(b)} {+} \Omega^{sm (b)} D^l U_{m(d)} \right) \right] \,.
\label{2CU319}
\end{equation}
The Lagrange multiplier can be found as
\begin{equation}
\lambda =  U_m^{(a)} \left[D_i {\cal J}^{im}_{(a)}
- {\cal I}^m_{(a)} \right]  \,,
\label{CU4}
\end{equation}
when the corresponding terms are already calculated.

\subsubsection{Equation for the axion field}

Variation of the action functional with respect to the pseudoscalar field $\phi$ gives the equation
\begin{equation}
\nabla^k \nabla_k \phi + m^2_A \sin{\phi}= - \frac{1}{4\Psi^2_0} \cos{\phi} {\cal L}^{\prime}({\cal I}) {}^*\! F^{(a)}_{mn} F^{mn}_{(a)} + {\cal G} \,,
\label{phi1}
\end{equation}
where the first term in the right-hand side describes the source related to the gauge field, and the pseudoscalar ${\cal G}$ given by
\begin{equation}
{\cal G} = \frac{1}{2\kappa \Psi^2_0} \left\{ \sin{\phi} \left[C_1 D_n U^{(a)}_m  D^n U_{(a)}^m  {+} C_2 D^m U^{(a)}_m  D_n U_{(a)}^n  {+}
C_3 D_n U^{(a)}_m D^m U_{(a)}^n  {+} \right. \right.
\label{phi2}
\end{equation}
$$
\left. \left. +C_4 U^p_{(a)}D_p U^{(a)}_m  U^q_{(b)}D_q U^{(b)m} \right] - \right.
$$
$$
\left. {-}2 \cos{\phi}\left[C_5 \Omega^{pq}_{(a)} D_pU_q^{(a)} {+} 2C_6 \Omega^{qm(b)} D_pU_m^{(a)}\left(U_{q(a)}  U^p_{(b)} {+} U_{q(b)} U^p_{(a)} \right) \right]\right\}
$$
presents the source associated with the color aether contribution.

\subsubsection{Equations for the gravitational field}

The last variation procedure in our context relates to the variation with respect to the metric.
The gravitational field is obtained to be described by the following set of equations:
\begin{equation}
 R_{pq} {-} \frac{1}{2} R  g_{pq} -  \Lambda g_{pq} = \lambda U^{(a)}_p  U_{(a)q}  +
 T^{(\rm U)}_{pq} + \kappa T^{(\rm YM)}_{pq} + \kappa T^{(\rm A)}_{pq} \,.
\label{CE1}
\end{equation}
The effective stress-energy tensor of the color vector fields $T^{(\rm U)}_{pq}$ is of the following form:
\begin{equation}
T^{(\rm U)}_{pq} = \frac12 g_{pq} {\cal K}^{ijmn}_{(a)(b)}  D_i U^{(a)}_m D_j U^{(b)}_n {+} D^m \left[U_{(a)(p}{\cal J}_{q)m}^{(a)} {-}
{\cal J}_{m(p}^{(a)}U_{q)(a)} {-}
{\cal J}_{(pq)}^{(a)} U_{(a)m} \right] {+}
\label{CE2}
\end{equation}
$$
+ \cos{\phi} \left[C_1\left(D_m U_p^{(a)} D^m U_{q(a)} - D_p U^{m(a)} D_q U_{m(a)} \right) +
C_4 U^m_{(a)} D_m U^{(a)}_p U^n_{(b)} D_n U^{(b)}_q \right] +
$$
$$
+ 4C_5 \sin{\phi} \left[\frac14 g_{pq}D_j U_{n(a)} \Omega^{jn(a)} - D_j U^{(a)}_{(p} \Omega^j_{q)(a)} \right] +
$$
$$
+ 4C_6 \sin{\phi} \left[2 D^j U^m_{(a)} U^{((b)}_j  U^{(a))}_{(p} \Omega_{q)m(b)}  + 2 D^j U_{(a)(p} \Omega^{m}_{q)(b)} U_{m}^{((a)} U_j^{(b))}+
 \right.
$$
$$
\left. +  D_i U^{m(a)} \epsilon_{jsm(p} D^j U^{(b)}_{q)} U^i_{((b)}U^s_{(a))} - g_{pq} D_j U^{m(b)} \Omega_{sm}^{(a)} U^{(j}_{(a)}  U^{s)}_{(b)} \right] \,.
$$
Here the parentheses $(pq)$ denote the symmetrization with respect to the coordinate indices $p$ and $q$, i.e., $a_{(p}b_{q)} = \frac12 (a_{p}b_{q}+a_{q}b_{p})$. The symbol $T^{(\rm YM)}_{pq}$ relates to the effective stress-energy tensor of the nonlinear Yang-Mills field
\begin{equation}
T^{(\rm YM)}_{pq} = {\cal L}^{\prime}({\cal I}) \left[\frac14 g_{pq}  F^{(a)}_{mn} F_{(a)}^{mn} -  F^{(a)}_{pm} F_{(a) qn} g^{mn}\right] +
g_{pq} \left[{\cal L} - {\cal I}{\cal L}^{\prime}({\cal I})\right]
\,.
\label{CE4}
\end{equation}
The last term in (\ref{CE1})
\begin{equation}
T^{(\rm A)}_{pq} =  \Psi^2_0 \left[\nabla_p \phi \nabla_q \phi + \frac12 g_{pq} \left(V(\phi) - \nabla_n \phi \nabla \phi^n \right)  \right]
\label{phi17}
\end{equation}
describes the canonic effective stress-energy tensor of the pure axion field.

\subsubsection{Short summary}

In this Section we established the new self-consistent SU(N) symmetric model of nonlinear interaction between gauge, vector, pseudoscalar and gravitational fields. The set of the model master equations is derived for arbitrary spacetime platform. This model as a whole deserves detailed analysis and has a lot of potential applications. But what we plan to do now? Our purpose is to make the first step and to analyze the solutions to the equation for the axion field (\ref{phi1}) with the sources, which correspond to the decay of the color aether in the early Universe. We assume that this story can be described in the framework of Bianchi-I spacetime model.

\section{Decay of the color aether}

\subsection{Ansatz about parallel fields}

In the work \cite{color2} we presented our version of the model of the color aether decay,  and of the transformation of the color aether into the canonic dynamic aether. The main idea is that a spontaneous color polarization took place, and the fields $U^{j(a)}$ happened to be parallel to one another in the group space. The idea of such parallelization is well known (see, e.g., \cite{Y,D,Z}). Mathematically, we have $U^{j(a)}= q^{a} U^j$, where $U^j$ is the timelike unit vector field associated with the velocity four-vector of the dynamic aether \cite{J1}. We assume that $U_kU^k=1$, and thus the constant color vector  $q^{(a)}$ is also unit, $G_{(a)(b)}q^{(a)}q^{(b)}=1$. At the decay moment the tensor of color polarization  $H^{(a)(b)} \equiv U^{(a)}_m U^{(b)}_n g^{mn}$ spontaneously transforms into
$H^{(a)(b)} \equiv q^{(a)} q^{(b)}$ with the diagonal structure (see \cite{color2} for details). Now we assume that in the epoch preceding the decay the Yang-Mills fields were generated just by the SU(N) symmetric aether, and thus, the decay of the color aether was accompanied by the decay of the gauge field. In other words, our second ansatz is that the multiplet of gauge field potentials $A^{(a)}_k$ also becomes parallel in the group space, i.e., $A^{(a)}_k = Q^{(a)} A_k$, and $Q^{(a)}$ is parallel to $q^{(a)}$, i.e., $Q^{(a)} {=} \nu q^{(a)}$ with some constant $\nu$.
With the proposed ansatz the master equations of the model simplify essentially.

\subsubsection{Reduced equations for the gauge fields}

First of all, the Yang-Mills fields become quasi-Maxwellian, i.e.,
\begin{equation}
F^{(a)}_{mn} = \nu q^{(a)}F_{mn} \,, \quad F_{mn} = \nabla_m A_n - \nabla_n A_m \,.
\label{D1}
\end{equation}
The terms $D_k U^{j(a)}$ and $\Omega^{pq(a)}$ convert, respectively, into
\begin{equation}
D_k U^{j(a)}=  q^{(a)}\nabla_k U^j \,, \quad \Omega^{pq(a)} = q^{(a)} \Omega^{pq} \,, \quad \Omega^{pq} = \frac12 \epsilon^{pqmn} \nabla_m U_n \,.
\label{D2}
\end{equation}
The extended Yang-Mills equations (\ref{0Col1}) take now the Maxwell-type form
\begin{equation}\label{06Col1}
\nabla_k \left[{\cal L}^{\prime}({\cal I}) \left(F^{ik} + \sin{\phi} \ {}^*\! F^{ik} \right) \right] = 0 \,,
\end{equation}
and the equation (\ref{dual002}) reads
\begin{equation}
\nabla_k {}^*\! F^{ik} =0 \,.
\label{dual52}
\end{equation}

\subsubsection{Reduced equations for the axion field}

With the proposed ansatz we can transform the axion field equation (\ref{phi1}) as follows
\begin{equation}
\nabla^k \nabla_k \phi {+} m^2_A \sin{\phi}= {-} \frac{\nu^2 \cos{\phi}}{4\Psi^2_0}  \ {\cal L}^{\prime}({\cal I}) \ {}^*\! F_{mn} F^{mn} {+} \frac{\cos{\phi}}{\kappa \Psi^2_0}\left[ 2(C_5{+}2C_6)  \ \omega^p DU_p  {-} C_5 \omega^{pq} \omega^{*}_{pq}\right] {+}
\label{D05}
\end{equation}
$$
+\frac{\sin{\phi}}{2\kappa \Psi^2_0}  \left[(C_1{+}C_3)\sigma_{mn}\sigma^{mn} + \frac13 \Theta^2 (C_1 {+}3C_2{+}C_3) {+}
(C_1{-}C_3) \omega_{mn}\omega^{mn}  {+} (C_1{+}C_4) {\cal D} U_m  {\cal D} U^m \right]  \,.
$$
Here we use the notations standard for the Einstein-aether theory, namely, the convective derivative ${\cal D}$, the scalar of the medium flow expansion $\Theta$, the projector $\Delta^p_m $, the skew-symmetric vortex tensor $\omega_{mn}$ and its dual $\omega^{* pq}$, the symmetric traceless shear tensor $\sigma_{mn}$, and the (pseudo) four-vector of the rotation velocity $\omega^p$:
\begin{equation}
{\cal D} \equiv U^p \nabla_p  \,, \quad \Theta \equiv \nabla_p U^p \,, \quad \omega_{mn} \equiv \frac12 \Delta^p_m  \Delta^q_n (\nabla_p U_q - \nabla_q U_p) \,,
\quad \Delta^p_m \equiv \delta^p_m - U^pU_m \,,
\label{D6}
\end{equation}
$$
\sigma_{mn} \equiv  \frac12 \Delta^p_m  \Delta^q_n (\nabla_p U_q + \nabla_q U_p) - \frac13 \Delta_{mn} \Theta \,, \quad \omega^{* pq} \equiv \frac12 \epsilon^{pqmn}\omega_{mn} \,, \quad
\omega^p \equiv \omega^{* pq} U_q  = \Omega^{pq}U_q  \equiv \Omega^p \,.
$$
The equation (\ref{D05}) admits the equilibrium-type  solutions $\phi = 2\pi n$ ($n$ is an integer), introduced and advocated in \cite{E1,E2,E3,E4}, if and only if
\begin{equation}
{\cal L}^{\prime}({\cal I})  {}^*\! F_{mn} F^{mn} = \frac{4}{\kappa \nu^2} \left[ 2(C_5{+}2C_6)  \ \omega^p DU_p  {-} C_5 \omega^{pq} \omega^{*}_{pq}\right]\,,
\label{D96}
\end{equation}
otherwise, the axion production takes place inevitably in the course of the color aether decay.

\subsubsection{Reduced equation for the vector field}

Since now ${\cal J}^{im}_{(a)} = q_{(a)}{\cal J}^{im}$ and ${\cal I}^{m}_{(a)} = q_{(a)}{\cal I}^{m}$, the balance equations (\ref{CU1}) take the form
\begin{equation}
\nabla_i {\cal J}^{im}  = \lambda \ U^m +  {\cal I}^{m} \,,
\label{CU19}
\end{equation}
where the corresponding reduced quantities are
\begin{equation}
{\cal J}^{im} = \cos{\phi}\left[C_1 \nabla^i U^{m} {+} C_2 g^{im} \Theta {+} C_3 \nabla^m U^{i} + C_{4} U^{i} {\cal D}U^{m} \right] +
\label{CU29}
\end{equation}
$$
+ \sin{\phi} \left[(2C_6-C_5)U_s \epsilon^{imsn}{\cal D}U_n - 8C_6 U^i \Omega^m  - 2C_5 \epsilon^{impq} \Omega_{pq}\right] \,,
$$
\begin{equation}
{\cal I}^l  =  C_{4} \cos{\phi} \nabla^l U^{n} \  {\cal D} U_{n}  + 4C_6  \sin{\phi} \left[\Omega^{lm} DU_m  - \Omega^m \nabla^l U_m  \right] \,.
\label{CU3192}
\end{equation}
The new coupling constant $C_5$ happens to be hidden in (\ref{CU319}).

\subsubsection{Modifications in the equations for the gravitational field}

Using the ansatz of parallel fields we have to replace $\lambda U^{(a)}_p  U_{(a)q}$ with $\lambda U_p U_q$ in (\ref{CE1}), and to delete the group indices $(a)$ in the stress-energy tensor of the Yang-Mills field (\ref{CE4}).
The stress-energy tensor (\ref{CE2}) converts now into
\begin{equation}
T^{(\rm U)}_{pq} =   \nabla^m \left[U_{(p}{\cal J}_{q)m} {-}
{\cal J}_{m(p}U_{q)} {-}
{\cal J}_{(pq)} U_{m} \right] {+}
\label{CE29}
\end{equation}
$$
+ \cos{\phi} \left[C_1\left(\nabla_m U_p \nabla^m U_{q} - \nabla_p U^{m} \nabla_q U_{m} \right) +
C_4 {\cal D} U_p {\cal D} U_q \right] +
$$
$$
+ \frac12 g_{pq} \cos{\phi} \left[C_1 \nabla^j U^m \nabla_j U_m + C_2 \Theta^2 + C_3 \nabla_j U_m  \nabla^m U^j + C_4 {\cal D}U^m {\cal D}U_m  \right] -
$$
$$
- 2C_5 \sin{\phi} \nabla_j U_{(p} \epsilon^j_{\ q)mn}\nabla^{m}U^n  +
$$
$$
+ 4C_6 \sin{\phi} \left[{\cal D}U^m U_{(p} \epsilon_{q)mls} \nabla^l U^s  + {\cal D}U_{(p} \epsilon_{q)mls} U^m \nabla^l U^s +
{\cal D}U^{m} U^s \epsilon_{jsm(p} \nabla^j U_{q)}   \right] \,.
$$
Next simplification of the total set of master equations appears, when we choose the spacetime platform, which corresponds to the homogeneous Bianchi-I cosmological model.

\subsection{Bianchi-I spacetime platform}

Below we consider the class of homogeneous spacetimes with the metric
\begin{equation}
ds^2 = dt^2 - a^2(t)dx^2 - b^2(t)dy^2 - c^2(t)dz^2  \,.
\label{Bianchi}
\end{equation}
We assume that all the unknown state functions inherit the spacetime symmetry and depend on the cosmological time only.
It is reasonable to choose the global unit timelike vector $U^i$ in the  form $U^i=\delta^i_t$ and to check that the master equations for this vector are consistent. It is well known that for the metric (\ref{Bianchi}) the covariant derivative is symmetric
\begin{equation}
\nabla_m U_n = \nabla_n U_m = \frac12 \dot{g}_{mn}   \,,
\label{B1}
\end{equation}
the vorticity tensor vanishes, $\omega_{mn}=0$, the acceleration  four-vector also is vanishing, ${\cal D}U^m =0$, and the scalar of expansion $\Theta$ is of the form
\begin{equation}
\Theta(t) \equiv \nabla_k U^k = \frac{\dot{a}}{a} + \frac{\dot{b}}{b} +\frac{\dot{c}}{c} \,.
\label{B2}
\end{equation}
As usual, the dot denotes the ordinary derivative with respect to time.

Additional simplification is connected with the constraint, obtained due to observations of the events, encoded as GW170817 and GRB 170817A. Based on the observation of the binary neutron star merger \cite{GRB17}, it was obtained that the ratio of the velocities of the gravitational and electromagnetic waves in vacuum satisfy the inequality $1-3 \times 10^{-15}< \frac{v_{\rm gw}}{c}<1+ 7 \times 10^{-16} $). The square of the velocity of the tensorial aether mode, calculated in \cite{J2}, is equal to $S^2_{(2)}= \frac{1}{1{-}(C_1{+}C_3)}$, thus  the sum of the parameters $C_1{+}C_3$ can be estimated as  $-6 \times 10^{-15}<C_1{+}C_3< 1.4 \times 10^{-15}$. Based on this fact we can consider that $C_1{+}C_3=0$ with very high precision, and put $C_3=-C_1$ in the formulas below.

\subsubsection{Solution to the equation for the gauge field}

Taking into account the symmetry of the reduced model, we can state that the equations (\ref{dual52}) admit the solution $F_{12}=const$, and the equations (\ref{06Col1}) can be reduced to one equation
 \begin{equation}\label{D38}
\frac{d}{dt}\left[{\cal L}^{\prime}({\cal I})\left(abc  F^{30} - F_{12} \sin{\phi}\right) \right] =0 \,.
\end{equation}
The solution to (\ref{D38}) is
\begin{equation}\label{D37}
F^{30}(t)   = F^{30}(t_0) \frac{a(t_0)b(t_0)c(t_0)}{a(t)b(t)c(t)} \cdot \frac{{\cal L}^{\prime}({\cal I})(t_0)}{{\cal L}^{\prime}({\cal I})(t)} + \frac{ F_{12}}{a(t)b(t)c(t)} \left[\sin{\phi(t)} -\sin{\phi(t_0)}\cdot \frac{{\cal L}^{\prime}({\cal I})(t_0)}{{\cal L}^{\prime}({\cal I})(t)} \right] \,,
\end{equation}
where $t_0$ is some specific time moment. For illustration we assume the following initial data:  $F^{30}(t_0)=0$ and $\phi(t_0) = \pi n$, so that below we work with the exact solution
\begin{equation}\label{D36}
F^{30}(t)   =  \frac{ F_{12}}{abc} \sin{\phi(t)} \,,
\end{equation}
which does not include the nonlinear function ${\cal L}({\cal I})$. For this elegant solution the quadratic terms
\begin{equation}\label{D35}
 \frac14 F^{(a)}_{mn}F^{mn}_{(a)} =  \frac{\nu^2 F^2_{12}}{2a^2b^2} \cos^2{\phi}\,, \quad  \frac14 {}^*\! F^{(a)}_{mn} F^{mn}_{(a)} = \frac{\nu^2F^2_{12}}{a^2b^2} \sin{\phi} \,, \quad {\cal I} = \frac{\nu^2 F^2_{12}}{2a^2b^2}(1+ \sin^2{\phi})
\end{equation}
also are rather simple.

\subsubsection{Solution to the equation for the vector field}

When we deal with the Bianchi-I model, for which ${\cal D}U_m= 0$, $\omega_{mn}=0$, $\Omega^{pq}=0$, $C_1{+}C_3=0$, we see that (\ref{CU319}) gives  ${\cal I}^l=0$, and one obtains from (\ref{CU29}) that
\begin{equation}\label{Dn75}
{\cal J}^{im} = C_2  \Theta \cos{\phi} g^{im} \,.
\end{equation}
Thus the equations (\ref{CU19}) admit the solution in the form $U^i=\delta^i_0$. Indeed, three equations from four ones are satisfied automatically, and the last equation gives the Lagrange multiplier $\lambda$:
\begin{equation}\label{D75}
\lambda= C_2 \frac{d}{dt}\left(\Theta \cos{\phi} \right) \,.
\end{equation}
In other words, two subsets of the master equations, namely, the equations for the gauge and vector fields, happen to be solved analytically.

\subsubsection{Key equation for the axion field}

The equation for the axion field also is simplified essentially in the Bianchi-I model with $C_1{+}C_3{=}0$, and it takes now the form
\begin{equation}
\ddot{\phi} + \Theta \dot{\phi} + \sin{\phi}\left[m^2_A +  \frac{\nu^2 F^2_{12}}{\Psi^2_0 a^2b^2} \cos{\phi} \ {\cal L}^{\prime}({\cal I}) - \frac{C_2}{2\kappa \Psi^2_0}  \Theta^2   \right]= 0   \,.
\label{D555}
\end{equation}
In this equation the guiding functions $a(t)$, $b(t)$, $\Theta(t)$ have to be found from the gravity field equations.

\subsubsection{Evolutionary equations for the gravitational field }

For the Bianchi-I model we obtain the simplified gravity field equations in the form
$$
\frac{\dot{a}}{a} \frac{\dot{b}}{b} + \frac{\dot{a}}{a} \frac{\dot{c}}{c} + \frac{\dot{b}}{b} \frac{\dot{c}}{c} = \Lambda - \frac12 C_2 \Theta^2 \cos{\phi} +
\frac12 \Psi^2_0 \left[V(\phi) + {\dot{\phi}}^2 \right] + \kappa {\cal L}(\cal I)\,,
$$
$$
\frac{\ddot{b}}{b} +  \frac{\ddot{c}}{c} + \frac{\dot{b}}{b} \frac{\dot{c}}{c} = \Lambda - C_2 \left(\Theta \cos{\phi} \right)^{\cdot} - \frac12 C_2 \Theta^2 \cos{\phi} +
\frac12 \Psi^2_0 \left[V(\phi) - {\dot{\phi}}^2 \right] +
\kappa  \left[{\cal L}({\cal I}) - 2 {\cal I} {\cal L}^{\prime}({\cal I}) \right] \,,
$$
$$
\frac{\ddot{a}}{a} +  \frac{\ddot{c}}{c} + \frac{\dot{a}}{a} \frac{\dot{c}}{c} = \Lambda - C_2 \left(\Theta \cos{\phi} \right)^{\cdot} - \frac12 C_2 \Theta^2 \cos{\phi} +
\frac12 \Psi^2_0 \left[V(\phi) - {\dot{\phi}}^2 \right]+
\kappa  \left[{\cal L}({\cal I}) - 2 {\cal I} {\cal L}^{\prime}({\cal I}) \right] \,,
$$
\begin{equation}\label{Ein}
\frac{\ddot{b}}{b} +  \frac{\ddot{a}}{a} + \frac{\dot{b}}{b} \frac{\dot{a}}{a} = \Lambda - C_2 \left(\Theta \cos{\phi} \right)^{\cdot} - \frac12 C_2 \Theta^2 \cos{\phi} +
\frac12 \Psi^2_0 \left[V(\phi) - {\dot{\phi}}^2 \right]+ \kappa  {\cal L}(\cal I)  \,.
\end{equation}
When the aether is absent and thus  $C_2=0$, these equations formally look similar to the equations (31)-(34) from \cite{Non3}, although, now the dependence of the function ${\cal I}$ on the axion field $\phi$ is principally different. However, this similarity makes it easier for us the qualitative monitoring of the behavior of the gravity field, if we take into account the corresponding numerical results from \cite{Non3}.

If the Bianchi-I model is characterized by the so-called Local Invariance, i.e., when $a(t)=b(t)$, one can introduce the auxiliary function $H= \frac{\dot{a}}{a}$ and exclude $c(t)$ from consideration using the relationship  $\frac{\dot{c}}{c} = \Theta {-} 2H$. Then, keeping in mind that on the solutions of the master equations the conservation law converts into identity, we obtain only two independent key equations for the gravity field with Local Invariance; in particular, for the quasi-linear model with ${\cal L}({\cal I})= {\cal I}$ this pair of key equations reads
\begin{equation}\label{Ein301}
2 H \Theta - 3H^2 = \Lambda - \frac12 C_2 \Theta^2 \cos{\phi} +
\frac12 \Psi^2_0 \left[V(\phi) + {\dot{\phi}}^2 \right] + \frac{\nu^2 F^2_{12}}{2a^4}(1+\sin^2{\phi})\,,
\end{equation}
\begin{equation}\label{Ein302}
 2 \dot{H} + 3 H^2 = \Lambda - C_2 \left(\Theta \cos{\phi} \right)^{\cdot} - \frac12 C_2 \Theta^2 \cos{\phi} +
\frac12 \Psi^2_0 \left[V(\phi) - {\dot{\phi}}^2 \right]+ \frac{\nu^2 F^2_{12}}{2a^4}(1+\sin^2{\phi})  \,.
\end{equation}
There is one interesting consequence from these equations
\begin{equation}\label{Ein303}
\frac{d}{dt}\left(H + \frac12C_2 \Theta \cos{\phi} \right) = -3H \left(H-\frac13 \Theta \right) \,.
\end{equation}
In such a scheme we deal with three key equations (\ref{D555}), (\ref{Ein301}) and, say, (\ref{Ein303}) for three unknown functions $\phi$, $H$ and $\Theta$.

\section{On the solutions to the equation for the axion field}

\subsection{Special solutions with $\phi = 2\pi n$: axions are in the equilibrium state}

When $\phi=2\pi n$, the equation  (\ref{D555}) is satisfied identically for arbitrary functions $a(t)$, $b(t)$, $\Theta(t)$, ${\cal L}(\cal I)$. For the values $\phi=2\pi n$ the potential of the axion field (\ref{14}) and its first derivative are equal to zero, $V(2\pi n)=0$, $V^{\prime}(2\pi n)=0$, i.e., the axion configuration is in the equilibrium state according to the terminology of the paper \cite{E2}. This state is stable. We can choose the number $n$ to be arbitrarily large, and such a solution can be considered as a final stable state of the axion field evolution.

The gravity field equations also become simplified. As an illustration, we consider the equations (\ref{Ein301}),(\ref{Ein303}) for this case:
\begin{equation}\label{0Ein301}
2 H \Theta - 3H^2 = \Lambda - \frac12 C_2 \Theta^2  + \frac{\nu^2 F^2_{12}}{2a^4}\,,
\end{equation}
\begin{equation}\label{Ein0303}
\frac{d}{dt}\left(H + \frac12C_2 \Theta  \right) = -3H \left(H-\frac13 \Theta \right) \,.
\end{equation}
In the asymptotic regime, when $\Lambda \neq 0$, $C_2 \neq - \frac23$ and $a \to \infty$, we obtain that
\begin{equation}\label{LL1}
H \to H_{\infty} = \frac13 \Theta_{\infty} \,, \quad \Theta_{\infty} = \sqrt{\frac{6 \Lambda}{2+3C_2} \,.}
\end{equation}
This asymptotic regime corresponds to isotropization of the Universe, and the behavior of the scale factors corresponds to the de Sitter type asymptote $\propto e^{H_{\infty} t} $. Typical method of solution of this system is the following: we find $\Theta$ from the quadratic equation (\ref{0Ein301})
\begin{equation}\label{LL2}
\Theta = - \frac{2}{C_2} H \pm \sqrt{\frac{4H^2}{C_2^2} + \frac{2}{C_2}  \left(\Lambda + 3H^2 + \frac{B_0^2}{x^4} \right)} \,,
\end{equation}
put it into (\ref{Ein0303}) and solve numerically the obtained nonlinear equation, keeping in mind that
\begin{equation}\label{LL3}
B_0^2\equiv \frac{\nu^2 F^2_{12}}{2a^4(t_0)} \,, \quad x \equiv \frac{a(t)}{a(t_0)} \,, \quad \frac{d}{dt} = xH(x) \frac{d}{dx} \,.
\end{equation}
Here we restrict ourselves by qualitative comments only.

When $\Lambda =0$, one can find the exact solutions to the system under discussion in the form
\begin{equation}\label{LL4}
H(x)= \frac{\beta(1+C_2)}{x^2} \,, \quad \Theta(x) =  \frac{\beta}{x^2} \,, \quad \beta = \frac{4 \sqrt3 B_0}{\sqrt{1-(12 C_2 +7)^2}} \,.
\end{equation}
The solution is real, when the Jacobson's parameter satisfies the condition
$-\frac23 <C_2<-\frac12$. The scale factors $a(t)$ and $c(t)$, related to these $H(x)$ and $\Theta(x)$, have form
\begin{equation}\label{4LL4}
\frac{a(t)}{a(t_0)} = \left[1+ 2\beta (1+C_2) (t-t_0)\right]^{\frac12} \,, \quad \frac{c(t)}{c(t_0)} = [1+ 2\beta (1+C_2) (t-t_0)]^{- \frac{2C_2+1}{2C_2+2}} \,.
\end{equation}
This regime can not provide the asymptotic isotropization, i.e., this submodel is appropriate for the early Universe only.

\subsection{Special solutions with $\phi = \pi (2n+1)$: axions are in the instable state}

When $\phi=\pi (2n+1)$, the equation  (\ref{D555}) again is satisfied identically, but now we see that $V(\pi (2n+1))=4m^2_A$, $V^{\prime}(\pi (2n+1))=0$, i.e., the state of the axion configuration relates to the maximum of the axion potential. This state is instable. Again we can choose the number $n$ to be arbitrarily. The key equations for the gravity field happen to be modified as follows: first, we have to replace in (\ref{0Ein301}) and (\ref{Ein0303}) the cosmological constant $\Lambda \to \Lambda_*  \equiv \Lambda + 2\Psi^2_0 m^2_A$; second, since $\cos{\pi (2n+1)}=-1$, we can simply change the sign in front of $C_2$. The results of the previous  subsection can be modified respectively.

\subsection{Dynamics of the axion field growth}

Let us focus on the qualitative analysis of the  equation (\ref{D555}). Formally speaking, it can be indicated as a modified equation of physical pendulum.

We would like to recall, that for the canonic physical pendulum the master equation $\ddot{x} {+} \omega_0^2 \sin{x}{=}0$ can be rewritten in the equivalent form $\dot{x}=y$, $\dot{y} = - \omega_0^2 \sin{\phi}$. The phase portrait of this autonomous two-dimensional dynamic system has the infinite number of critical stationary points of two types: the saddle points at $y=0$, $x= \pm (2m+1) \pi$, and the centers at $y=0$, $x= \pm 2m \pi $. The saddle points are linked by separatrices, thus dividing the phase portrait into zones of finite and infinite motion (see \cite{Arnold} for details).   In our context it is important to stress that, when we consider the phase trajectory with noncritical initial data $y_0 \neq 0$ for the starting points of both types: $x_0= \pm (2m+1) \pi$, and $x_0= \pm 2m \pi$, we can choose the value $y_0$ so that the pendulum motion will be infinite, and the coordinate $x$ will grow with time.

Equation (\ref{D555}) also can be presented in terms of two-dimensional dynamic system
\begin{equation}
\dot{\phi} = y \,, \quad \dot{y} = - \Theta y -  {\cal H} \sin{\phi}  \,,
\label{D888}
\end{equation}
where
\begin{equation}
{\cal H}(t,\phi) = \left[m^2_A +  \frac{\nu^2 F^2_{12}}{\Psi^2_0 a^2(t)b^2(t)} \cos{\phi} \ {\cal L}^{\prime}({\cal I}(t,\phi)) - \frac{C_2}{2\kappa \Psi^2_0}  \Theta^2(t) \right] \,.
\label{D88}
\end{equation}
Again, the points $y=0$, $\phi = \pi n$ are critical and stationary, they do not move on the phase plane. But there are a few novel details. First of all, this dynamic system is not autonomous, and its qualitative analysis is possible in the time intervals, in which the functions $a(t)$, $b(t)$, $\Theta(t)$ change slowly, so that in these time intervals they can be approximately considered as constant. In this scheme the type of critical points $y=0$, $\phi = \pi n$ depends on the sign of the term ${\cal H}(t,\phi)$ on the fixed time interval. Second important detail is connected with possible new critical (nonstationary) points, which can appear, drift and disappear during different time intervals.

\subsubsection{The first case: the function ${\cal H}$ has no zeros and ${\cal H}>0$  }

The function ${\cal H}$ is positive on the whole admissible  time interval, when
\begin{equation}
\cos{\phi} \ {\cal L}^{\prime}({\cal I}) > \frac{\Psi^2_0 a^2 b^2}{\nu^2 F^2_{12}}\left[ \frac{C_2}{2\kappa \Psi^2_0}  \Theta^2 -m^2_A  \right] \,.
\label{D77}
\end{equation}
In this situation the critical points $y=0$, $\phi= \pm (2m+1) \pi$ and $y=0$, $\phi = \pm 2m \pi $ remain the saddle points and the centers, respectively, and the instantaneous phase portrait of the dynamic system looks like the one of the canonic physical pendulum.  To illustrate the idea, we assume that $C_2=0$ and ${\cal L} = {\cal I}$. Then the simplified requirement (\ref{D77})
\begin{equation}
\cos{\phi} > - \frac{m^2_A \Psi^2_0 a^2 b^2}{\nu^2 F^2_{12}}
\label{D44}
\end{equation}
is satisfied, when  $m^2_A \Psi^2_0 > \frac{\nu^2 F^2_{12}}{a^2(t) b^2(t)}$.

\subsubsection{The second case: the function ${\cal H}$ has no zeros and ${\cal H}<0$}

The function ${\cal H}$ is negative on the whole admissible  time interval, when
\begin{equation}
\cos{\phi} \ {\cal L}^{\prime}({\cal I}) < \frac{\Psi^2_0 a^2 b^2}{\nu^2 F^2_{12}}\left[ \frac{C_2}{2\kappa \Psi^2_0}  \Theta^2 -m^2_A  \right] \,.
\label{D33}
\end{equation}
For this situation
the critical points $y=0$, $\phi= \pm (2m+1) \pi$ become centers, and $y=0$, $\phi = \pm 2m \pi $ convert into the saddle points. The phase portrait seems to be the one for the physical pendulum shifted by the value $\pi$ along the horizontal axis. By the way, when $C_2=0$, ${\cal L} = {\cal I}$, this is not possible.

\subsubsection{The third case: the function ${\cal H}$ has zeros and changes the sign}

If the solutions to the equation ${\cal H}=0$ exist, and we deal with the trigonometric equation
\begin{equation}
\cos{\phi}  = \frac{\Psi^2_0 a^2 b^2}{\nu^2 F^2_{12}  {\cal L}^{\prime}({\cal I})}\left[ \frac{C_2}{2\kappa \Psi^2_0}  \Theta^2 -m^2_A  \right] \,,
\label{D22}
\end{equation}
one can write the roots of this equation as $\phi = \pm \phi_* + 2\pi k$, where the solution $\phi_*$ belongs to the interval $(0, \pi)$.
We obtain now a new series of critical points, $y=0$, $\phi = \pm \phi_* + 2\pi k$, if
\begin{equation}
\left| \frac{\Psi^2_0 a^2 b^2}{\nu^2 F^2_{12}  {\cal L}^{\prime}({\cal I})}\left( \frac{C_2}{2\kappa \Psi^2_0}  \Theta^2 -m^2_A  \right)\right| <1 \,.
\label{D22w}
\end{equation}
In particular, when $C_2=0$ and ${\cal L} = {\cal I}$ and the product $a(t)b(t)$ grows, this requirement reads
$m^2_A \Psi^2_0 < \frac{\nu^2 F^2_{12}}{a^2(t_0) b^2(t_0)}$, and it can be fulfilled for rather strong magnetic component of the gauge field.

What is the type of these new critical nonstationary points? As usual, we put $\phi = \pm \phi_* + \xi$ and $y=\eta$
with small $\xi$ and $\eta$, and obtain the linearized system
\begin{equation}
\dot{\xi} = \eta \,, \quad \dot{\eta} = - \Theta \eta + \xi \ \sin^2{\phi_*} \frac{\nu^2 F^2_{12} {\cal L}^{\prime}({\cal I})}{\Psi^2_0 a^2 b^2} \,.
\label{D221}
\end{equation}
The corresponding characteristic equation has the roots
\begin{equation}
\lambda_{1,2} = - \frac12 \Theta \pm \sqrt{\frac14 \Theta^2 + \sin^2{\phi_*} \frac{\nu^2 F^2_{12} {\cal L}^{\prime}({\cal I})}{\Psi^2_0 a^2 b^2} }\,.
\label{D223}
\end{equation}
Keeping in mind that in the quasi-linear case ${\cal L}^{\prime}({\cal I}) = 1 >0$, we assume that in general nonlinear case we can choose the models with ${\cal L}^{\prime}({\cal I}) >0$ only. With this assumption we see that the roots of the characteristic equations $\lambda_1$ and $\lambda_2$ are real and have opposite signs; thus, the new critical points happen to be the saddle points. On the phase portrait of the dynamic system the new saddle points appear between the old critical points (for instance, between $\phi=0$ and $\phi = \pi$, between $\phi=\pi$ and $\phi = 2\pi$, etc). This event reconstructs the phase portrait, since now all stationary points $\phi = \pi k$ become centers, and the number of finite zones doubles. When the product $a^2(t)b^2(t)$ grows, there comes such a moment in time, when the function ${\cal H}$ (\ref{D88}) losses the possibility to have zeros. The last possibility to have zeros takes place when $\cos{\phi}=1$ or $\cos{\phi}=-1$, depending on the values of guiding parameters. This means that at that time moment $+\phi_*= - \phi_* =0$ or $+\phi_*= - \phi_* =\pi$. On the phase portrait two new saddle points, which harbor the points $\phi = (2k+1)\pi$, are converging and finally coincide, thus removing additional zones of finite motion and recovering the old saddle points at $\phi = (2m+1)\pi$. Illustration of typical phase portrait is presented on Fig.1.

We have to stress, that in this case the zones of infinite motion exist, and the axion field can grow to an arbitrarily large value.

\begin{figure}[h]
\includegraphics[width=13 cm]{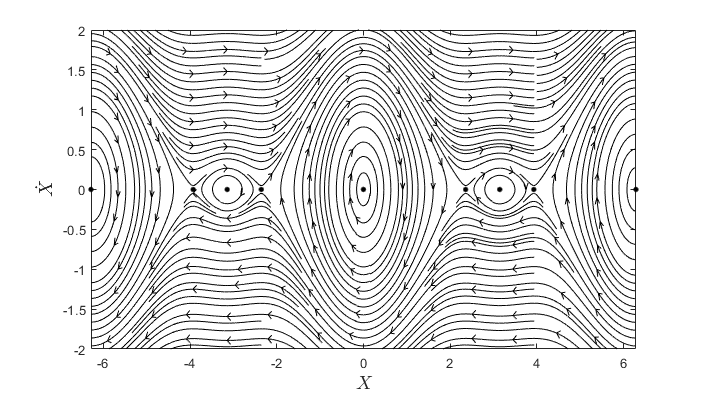}
\caption{Illustration of a typical phase portrait of the dynamic system (\ref{D888}) with (\ref{D88}); when the function ${\cal H}$ (\ref{D88}) has no zeros, the saddle points harboring the centers converge thus transforming such phase portrait to the one for the physical pendulum.  The phase portrait contains two zones of infinite motion, which are located above the upper line formed by separatrices and below the lower one; the corresponding curves describe infinite grow of the axion field. \label{fig1}}
\end{figure}

\section{Conclusions and outlook}

The main idea we can conclude from this work is that the decay of the color aether, which took place in the early Universe, can be accompanied by intensive axion production. Why do we think so?

The evolutionary equation for the axion field (\ref{D555}), which we obtained in the framework of the nonlinear version of the self-consistent SU(N) symmetric Einstein-Yang-Mills-aether-axion model, is, in fact, the generalized master equation of physical pendulum, and this equation admits the formal translational symmetry $\phi \to \phi + 2\pi n$ with integer $n$.  The typical instantaneous phase portrait of the corresponding non-autonomous dynamic system (\ref{D888}), depicted on Fig.1, shows that, as in the case of classical physical pendulum, the phase portrait contains two zones of infinite motion (they  are located above the upper line formed by separatrices and below the lower one). The axionic system, which starts to evolve with the initial value  $\phi(t_0) = \pi n$ (see the requirement transforming the solution (\ref{D37}) to (\ref{D36})) can enter the zones of infinite motion, if the initial value $\dot{\phi}(t_0)$ is appropriate. Clearly, when $\phi(t_0) = \pi$, i.e., the pendulum is in the inverted state, and the starting point is located on the phase plane near the saddle point, a small disturbance is enough to provide the infinite motion. For the starting point $\phi(t_0)=0$, the corresponding value $\dot{\phi}(t_0)$ should be big enough to provide the entrance to the zone of infinite motion.

 What does this means? In the described situation the pseudoscalar field $\phi$ grows quasi-periodically, and its final value can be estimated as $\phi = 2\pi n_{\infty}$. The final value $n_{\infty}$ is predetermined by the behavior of the function $\Theta(t)$ in the related model of the spacetime evolution. In this sense it is interesting to analyze the term ${\cal H}$ (\ref{D88}) in different cosmological epochs. When the term $a^2b^2$ is small and $F^2_{12}$ is big, the second term in ${\cal H}$ is the leading one, so that the typical time scale for the axion field oscillations $T$ relates to  $T \to \frac{\psi_0 ab}{\nu |F_{12}| \sqrt{{\cal L}^{\prime}({\cal I})}}$. Thus, in the early Universe the typical time scale is predetermined by the gauge field strength, and it is rather small. For big values of the function $a^2(t)b^2(t)$  the second term in (\ref{D88}) is negligible, and the typical time scale is predetermined by the axion interaction with the aether, $T \to \sqrt{\frac{2\kappa}{\left|C_2\right|}} \  \frac{\Psi_0}{\Theta}$, or by the axion mass, $T \to \frac{1}{m_A}$, if $C_2=0$. We agree that in this point some estimations are necessary, and we plan to consider them in a special letter.

 In order to formulate our further plans, we would like to make the following three comments.

 1. In the paper \cite{Non3}, using  the nonlinear axion electrodynamics, we have shown that the interaction between the axionic dark matter and global magnetic field can lead to the  anomalous flares of the axionically induced electric field, but the axion field itself remains in the finite zone of evolution. In this work we generalized the U(1) symmetric axion electrodynamics, and formulated the nonlinear SU(N) symmetric generalized axion chromodynamics. In contrast to the mentioned result of \cite{Non3}, we have shown that the axion field can grow to an arbitrarily large value, but the gauge field remains in the finite zone. Such a difference appeared, since in \cite{Non3} we used the internal Jackson's SO(2) symmetry of classical electrodynamics, but here we assumed that the whole SU(N) symmetric model can inherit the discrete symmetry of the physical pendulum.

 2. In the paper \cite{AA} we considered full-format generalization of the linear axion-aether theory, however, the electromagnetic field did not include into consideration. In the work \cite{BL} we studied the general formalism of the U(1) symmetric  Einstein-Maxwell-aether theory, but the axion field did not include into consideration. Now we made the step to finish the proposed program, namely, we considered the model with SU(N) symmetry instead of U(1), i.e., we introduced the Yang-Mills fields instead of the Maxwell field, the color aether instead of the dynamic aether, and considered the coupling of the axion field to both of them.

 3. In the papers \cite{E1,E2,Non2} we studied the direct influence of the dynamic aether on the axionic dark matter by using the axion field potential, which depends on the invariants attributed to the dynamic aether. Now we took into account the backreaction of the axion field on the color aether by using the generalization of the Jacobson's constitutive tensor.

 Putting together all these finds, we hope in the nearest future to build a catalog of models, which predict the axionic dark matter production on the different stage of the Universe evolution.

\vspace{5mm}
\noindent
{\bf Acknowledgments}

\noindent
The work was supported by the Russian Science Foundation (Grant No 21-12-00130). The authors are grateful to Dr. Bochkarev V.V. for valuable comments.

\end{document}